# Visual response properties of MSTd emerge from a sparse population code

Michael Beyeler[1], Nikil Dutt[1,2], and Jeffrey L. Krichmar[1,2]

[1]Department of Computer Science, University of California Irvine, Irvine, CA, USA

[2]Department of Cognitive Sciences, University of California Irvine, Irvine, CA, USA

**Abstract**

Neurons in the dorsal subregion of the medial superior temporal (MSTd) area respond to large, complex patterns of retinal flow, implying a role in the analysis of self-motion. Some neurons are selective for the expanding radial motion that occurs as an observer moves through the environment ("heading"), and computational models can account for this finding. However, ample evidence suggests that MSTd neurons may exhibit a continuum of visual response selectivity to large-field motion stimuli, but the underlying computational principles by which these response properties are derived remain poorly understood. Here we describe a computational model of MSTd based on the hypothesis that neurons in MSTd efficiently encode the continuum of large-field retinal flow patterns on the basis of inputs received from neurons in MT, with receptive fields that resemble basis vectors recovered with nonnegative matrix factorization (NMF). These assumptions are sufficient to quantitatively simulate neurophysiological response properties of MSTd cells such as radial, circular, and spiral motion tuning, suggesting that these properties might simply be a by-product of MSTd neurons performing dimensionality reduction on their inputs. At the population level, model MSTd accurately predicts heading using a sparse distributed code, consistent with the idea that biological MSTd might operate in a sparseness regime well-suited to efficiently encode a number of self-motion variables. The present work provides an alternative to the template-model view of MSTd, and offers a biologically plausible account of the receptive field structure across a wide range of visual response properties in MSTd.

## 1. Introduction

Neurons in the dorsal subregion of the medial superior temporal (MSTd) area of monkey extrastriate cortex respond to relatively large and complex patterns of retinal flow, often preferring a mixture of translational, rotational, and to a lesser degree deformational flow components (Saito et al., 1986; Tanaka and Saito, 1989; Duffy and Wurtz, 1991a, b; Orban et al., 1992; Graziano et al., 1994; Lagae et al., 1994; Mineault et al., 2012). This has led researchers to suggest that MSTd might play a key role in visual motion processing for self-movement perception. In fact, one of the most commonly documented response properties of MSTd neurons is that of heading selectivity (Tanaka and Saito, 1989; Duffy and Wurtz, 1995; Lappe et al., 1996; Britten and Van Wezel, 2002; Page and Duffy, 2003; Gu et al., 2006; Logan and Duffy, 2006; Gu et al., 2012), and computational models can account for this finding (Zhang et al., 1993; Perrone and Stone, 1994; Beintema and van den Berg, 1998; Perrone and Stone, 1998; Zemel and Sejnowski, 1998; Beintema and van den Berg, 2000). Ample evidence suggests that MSTd neurons also encode a number of visual motion-related variables (Bremmer et al., 1998; Ben Hamed et al., 2003; Brostek et al., 2014) and exhibit a continuum of visual response selectivity to large-field motion stimuli (Duffy and Wurtz, 1991a; Graziano et al., 1994; Lagae et al., 1994). However, little is known about the underlying computational principles by which these response properties are derived.

In this paper we describe a computational model of MSTd based on the hypothesis that neurons in MSTd efficiently encode the continuum of large-field retinal flow patterns encountered during self-movement on the basis of inputs received from neurons in MT. Nonnegative matrix factorization (NMF) (Paatero and Tapper, 1994; Lee and Seung, 1999, 2001), a linear dimensionality reduction technique, is used to find a set of nonnegative basis vectors, which are interpreted as MSTd-like receptive fields. NMF is similar to principal component analysis (PCA) and independent component analysis (ICA), but unique among these dimensionality reduction techniques in that it can recover representations that are often sparse and "parts-based", much like the intuitive notion of combining parts to form a whole (Lee and Seung, 1999).

Remarkably, applying NMF to MT-like patterns of activity was sufficient to recover a wide range of neurophysiologically recorded properties of MSTd cells such as a continuum of response selectivity to radial, circular, and spiral motion patterns, suggesting that these properties might simply be a by-product of MSTd neurons performing dimensionality reduction on their inputs. At the population level, a sparse, distributed code emerges (Olshausen and Field, 1996, 1997; Krekelberg et al., 2001; Ben Hamed et al., 2003), which enables model MSTd to efficiently and accurately encode heading. The model performs better at predicting heading than a template model with an equal number of heading templates (Perrone and Stone, 1994, 1998), despite the fact that the present MSTd model is not engineered to perform this task. Interestingly, we find that the sparseness regime in which the model achieves the lowest heading prediction error is also the regime in which recovered basis vectors most resemble receptive fields of macaque MSTd. This finding suggests that neurons in MSTd might operate according to the efficient-coding and free-energy principles (Attneave, 1954; Barlow, 1961; Linsker, 1990; Simoncelli and Olshausen, 2001; Friston et al., 2006; Friston, 2010), which would allow them to efficiently encode a number of self-motion related variables (Bremmer et al., 1998; Ben Hamed et al., 2003; Brostek et al., 2014). Thus the present work provides an alternative to the template model-view of MSTd (Perrone and Stone, 1994, 1998), and offers a biologically plausible account of the receptive field structure across a wide range of visual response properties in MSTd.

## 2. Methods

The overall architecture of the model is depicted in Figure 1. Visual input to the system encompassed a range of idealized 2D flow fields made from first-order motion components (translations, rotations, and deformations). We sampled the full range of first-order components and shifted the center of motion (COM) of these flow fields across spatial locations to yield a total of $S$ flow fields, comprising the input stimuli. Each flow field was processed by an array of $F$ MT-like motion sensors (MT-like model units), each tuned to a specific direction and speed of motion. The activity values of the MT-like model units were then arranged into the columns of an $F \times S$ matrix, **V**, which served as input for nonnegative matrix factorization (NMF) (Paatero and Tapper, 1994; Lee and Seung, 1999, 2001). NMF belongs to a class of dimensionality reduction methods that can be used to linearly decompose a multivariate data matrix, **V**, into an inner product of two reduced-rank matrices, **W** and **H**, such that $V \approx WH$. The first of these reduced-rank matrices, **W**, contains as its columns a total of $B$ nonnegative basis vectors of the decomposition. The second matrix, **H**, contains as its rows the contribution of each basis vector in the input vectors (the hidden coefficients). These two matrices are found by iteratively reducing the residual between **V** and **WH** using an alternating nonnegative least squares method. In our experiments, the only open parameter of the NMF algorithm was the number of basis vectors, $B$. We interpreted the columns of **W** as the weight vectors of a total of $B$ MSTd-like model units. Each weight vector had $F$ elements representing the synaptic weights from the array of MT-like model units onto a particular MSTd-like model unit. The response of an MSTd-like model unit was given as the dot product of the $F$ MT-like unit responses to a particular input stimulus and the corresponding nonnegative synaptic weight vector, **W**. Crucially, once the weight matrix **W** was found, the model remained fixed across all experiments. The following subsections explain the model in detail.

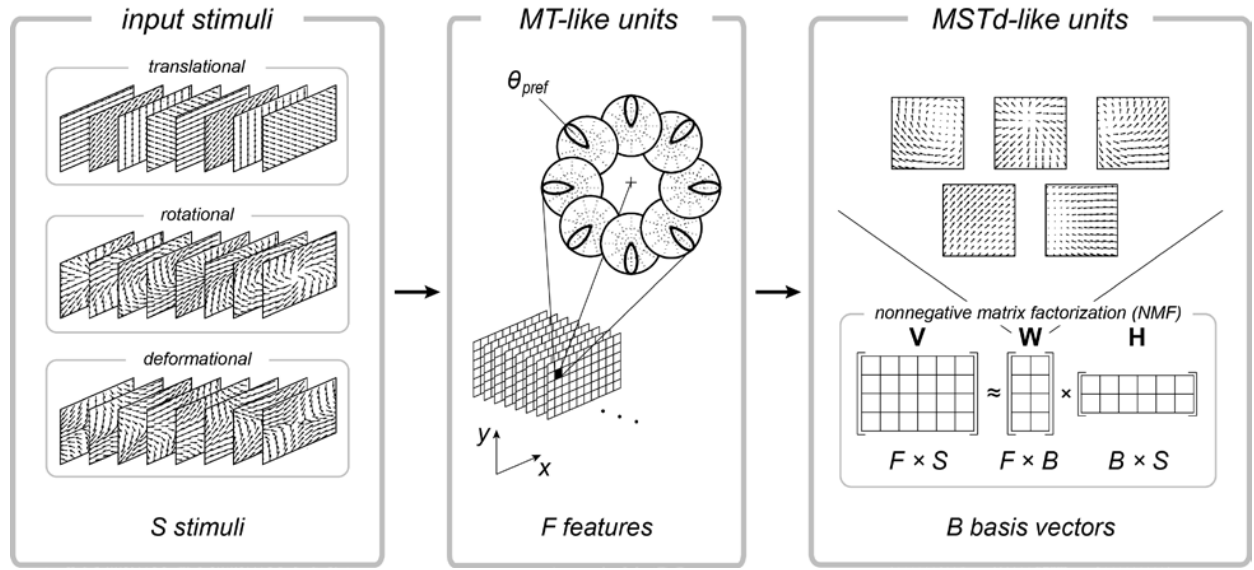

Figure 1: Overall model architecture. A number S of 2D flow fields made from first-order motion components were processed by an array of F MT-like motion sensors, each tuned to a specific direction of motion. MT-like activity values were then arranged into the columns of a data matrix, V, which served as input for nonnegative matrix factorization (NMF). The output of NMF were two reduced-rank matrices, W (containing B nonnegative basis vectors) and H (containing hidden coefficients). Columns of W (basis vectors) were then interpreted as weight vectors of MSTd-like model units.

## 2.1 Optic flow stimuli

Input to the model was a computer-generated $32 \times 32$ array of idealized optic flow vectors. The direction of flow at pixel location $(x, y)$ was given by an angle $\theta$:

$$\theta(x, y) = \sigma \operatorname{atan}\left(\frac{y - y_{\text{COM}}}{x - x_{\text{COM}}} + \phi\right) \bmod 2\pi, \tag{1}$$

where $(x_{\text{COM}}, y_{\text{COM}})$ were the coordinates of the COM, $\phi$ was an offset ranging between 0 and $2\pi$, and $\sigma = \{-1, 0, 1\}$. Setting $\sigma = 0$ and $\phi \in [0, 2\pi)$ generated all translational 2D flow fields. Setting $\sigma = 1$, $\phi \in [0, 2\pi)$, and $x_{\text{COM}}, y_{\text{COM}} \in [0, 31]$ generated all rotational 2D flow fields, which included radial flow fields (expansion, contraction), circular flow fields, and mixtures thereof (spirals). This range of flow fields has been termed "spiral space" in other publications (Graziano et al., 1994). Analogously, setting $\sigma = -1$, $\phi \in [0, 2\pi)$, and $x_{\text{COM}}, y_{\text{COM}} \in [0, 31]$ generated all deformational flow fields. There was no noise in the stimulus, and for the sake of simplicity we considered only a single speed of motion.

In order to reproduce population data comparable to MSTd, we sampled the offset, $\phi$, of all input stimuli according to the empirically observed likelihood of preferred directions of motion (Graziano et al., 1994; Clifford et al., 1999). This is akin to suggesting that the model samples input stimuli from spiral space with a bias towards expanding flow fields, as encountered during forward motion. Other models have made similar assumptions about sampling input stimuli (Perrone and Stone, 1998; Zemel and Sejnowski, 1998). The best fit with biological data was achieved with an input stimulus set consisting of 30% planar stimuli, 60% rotational stimuli, and 10% deformational stimuli.

## 2.2 MT stage

Each flow field was processed by an array of idealized MT-like model units, each selective to a particular direction of motion, $\theta_{\text{pref}}$, at a particular spatial location, $(x, y)$. The activity of each unit was given as a Gaussian function based on the difference between its preferred direction of motion and the direction of flow at that position:

$$r_{\text{MT}}(x, y; \theta_{\text{pref}}) = \exp\left(-\frac{\left(\theta(x, y) - \theta_{\text{pref}}\right)^2}{2\sigma_{\text{MT}}^2}\right), \tag{2}$$

where the bandwidth parameter was chosen to be $\sigma_{\text{MT}} = 0.\overline{3}$, so that the resulting tuning width (full width at half-maximum) was roughly 45°, which lies halfway in the 30° − 60° range of tuning widths reported for neurons in macaque MT (Rodman and Albright, 1989). At each spatial location there were a total of eight MT-like model units with different preferred directions of motion (in 45° increments), yielding a total of $F = 32 \times 32 \times 8 = 8192$ units. The activity pattern of these eight units per pixel thus acted as a population code for the local direction of motion.

The size and shape of receptive fields (RFs) in macaque MT change systematically with eccentricity, covering 31 deg$^2$ of the visual field near the fovea (which corresponds to a circular RF with ~3° radius), but approaching the size of MSTd RFs in the periphery (Raiguel et al., 1997). For the sake of simplicity, we did not model these effects. Rather, we assumed that all MT-like model units had a single-pixel (~2°) radius, which was slightly smaller than the average RF in macaque MT, but not uncommon (Raiguel et al., 1995).

## 2.3  Nonnegative matrix factorization (NMF)

We hypothesized that appropriate synaptic weights of the feed-forward projections from MT to MSTd could be learned with NMF (Paatero and Tapper, 1994; Lee and Seung, 1999, 2001). NMF belongs to a class of methods that can be used to decompose multivariate data into an inner product of two reduced-rank matrices, where one matrix contains nonnegative basis vectors and the other contains nonnegative activation vectors (hidden coefficients). The nonnegativity constraints of NMF enforce the combination of different basis vectors to be additive, leading to representations that are often parts-based and sparse (Lee and Seung, 1999). When applied to neural networks, these nonnegativity constraints correspond to the notion that neuronal firing rates are never negative and that synaptic weights are either excitatory or inhibitory, but they do not change sign (Lee and Seung, 1999).

Assume that we observe data in the form of a large number of i.i.d. random vectors, $\vec{v}^{(i)}$, such as the neuronal activity of a population of MT neurons in response to a visual stimulus, where $i$ is the sample index. When these vectors are arranged into the columns of a matrix **V**, linear decompositions describe these data as **V** ≈ **WH**, where **W** is a matrix that contains as its columns the basis vectors of the decomposition. The rows of **H** contain the corresponding hidden coefficients that give the contribution of each basis vector in the input vectors. Like principal component analysis (PCA), the goal of NMF is then to find a linear decomposition of the data matrix **V**, with the additional constraint that all elements of the matrices **W** and **H** be nonnegative. In contrast to independent component analysis (ICA), NMF does not make any assumptions about the statistical dependencies of **W** and **H**. The resulting decomposition is not exact; **WH** is a lower-rank approximation to **V**, and the difference between **WH** and **V** is termed the reconstruction error. Perfect accuracy is only possible when the number of basis vectors approaches infinity, but good approximations can usually be obtained with a reasonably small number of basis vectors (Pouget and Sejnowski, 1997).

We used the standard `nnmf` function provided by MATLAB R2014a (MathWorks, Inc.) which, using an alternating least-squares algorithm, aims to iteratively minimize the root-mean-squared residual $D$ between **V** and **WH**:

$$D = \frac{\|\mathbf{V} - \mathbf{WH}\|}{\sqrt{FS}}, \qquad (3)$$

where $F$ is the number of rows in **W** and $S$ is the number of columns in **H** (see Figure 1). **W** and **H** were normalized so that the rows of **H** had unit length. The output of NMF is not unique because of random initial conditions (i.e., random weight initialization).

The only open parameter was the number of basis vectors, $B$, whose value had to be determined empirically. In our simulations we examined a range of values ($B = 2^i$, where $i = \{2, 3, \ldots, 7\}$, see Section "Representation of heading using a sparse population code"), but found that $B = 16$ led to basis vectors that most resembled receptive fields found in macaque MSTd. In order to facilitate statistical comparisons between model responses and biological data, NMF with $B = 16$ was repeated four times with different random initial conditions to generate a total of $N = 64$ MSTd-like model units.

### 2.4 MSTd stage

We interpreted the resulting columns of **W** as the weight vectors from the population of $F$ MT-like model units onto a population of $B$ MST-like model units. The activity of the $b$-th MSTd-like unit, $r_{\text{MSTd}}^b$, was given as the dot product of all $F$ MT-like responses to a particular input stimulus, $i$, and the unit's corresponding nonnegative weight vector:

$$r_{\text{MSTd}}^b = \vec{v}^{(i)} \vec{w}^{(b)}, \tag{4}$$

where $\vec{v}^{(i)}$ was the $i$-th column of **V** and $\vec{w}^{(b)}$ was the $b$-th column of **W**. This is in agreement with the finding that MSTd response are approximately linear in terms of their feed-forward input from area MT (Tanaka et al., 1989; Duffy and Wurtz, 1991b), which provides one of the strongest projections to MST in the macaque (Boussaoud et al., 1990). In contrast to other models (Zemel and Sejnowski, 1998; Grossberg et al., 1999; Lappe, 2000; Mineault et al., 2012), no additional nonlinearities were required to fit the data presented in this paper.

RFs in MSTd have characteristics that are largely independent of eccentricity, being mostly circular or elliptic and covering $2170 \text{deg}^2$ of the visual field, which corresponds to a circular RF with ~26° radius, but their size and shape can vary considerably depending on the stimulus and the plotting criteria used (Raiguel et al., 1997). For the sake of simplicity, and because mapping pixels to degrees of visual field is arbitrary, we instead assumed that the RFs of all MSTd-like model units were centered on the same $32 \times 32$ pixel input, making them roughly an order of magnitude larger than the RFs of MT-like model units. If we assume, for the sake of argument, that a pixel roughly encompassed 2.5° of the visual field, then the visual stimulus encompassed a 80° × 80° area, the RF radius of MT-like model units roughly corresponded to 2° (which was slightly smaller than the average RF in macaque MT, but not uncommon), and the RF radius of MST-like model units corresponded to roughly 40° (which was slightly larger than the average RF in macaque MSTd, but not uncommon) (Duffy and Wurtz, 1991a, 1995; Raiguel et al., 1997).

The receptive fields of MSTd-like model units were visualized as flow fields by calculating the population vector, $\vec{v}_{\text{pop}}$ (Georgopoulos et al., 1982), across the synaptic weight contributions of MT-like model units coding for a particular spatial location $(x, y)$:

$$\vec{v}_{\text{pop}}(x, y) = \sum_{\theta} \begin{bmatrix} cos\theta \\ sin\theta \end{bmatrix} w(x, y, \theta), \qquad (5)$$

where $w(x, y, \theta)$ was the element of **W** that was associated with $r_{\text{MT}}(x, y; \theta)$ in (2), and the sum was over all directions of flow, $\theta$.

## 2.5 Sparseness

We computed a sparseness metric for the modeled MSTd population activity according to the definition of sparseness by Vinje and Gallant (2000):

$$s = \left(1 - \frac{1}{N}\frac{(\sum_i r_i)^2}{\sum_i r_i^2}\right) \Big/ \left(1 - \frac{1}{N}\right). \tag{6}$$

Here, $s \in [0,1]$ is a measure of sparseness for a signal $r$ with $N$ sample points, where $s = 1$ denotes maximum sparseness and is indicative of a local code, and $s = 0$ is indicative of a dense code. In order to measure how many MSTd-like model units were activated by any given stimulus (population sparseness), $r_i$ was the response of the $i$-th neuron to a particular stimulus and $N$ was the number of neurons. In order to determine how many stimuli any given model unit responded to (lifetime sparseness), $r_i$ was the response of a neuron to the $i$-th stimulus and $N$ was the number of stimuli. Population sparseness was averaged across stimuli and lifetime sparseness was averaged across neurons.

## 3. Results

### 3.1 Gaussian tuning in spiral space

The visual processing of MSTd has been summarized as template matching with radial, rotational, and translational motion, or combinations of these features (Orban et al., 1992; Graziano et al., 1994). That is, neurons in MSTd signal how well the flow present on the retina matches their preferred flow component or mixture of components, decreasing their response when increasing amounts of non-preferred components are added (Orban et al., 1992). Tanaka et al. (1989) concluded that the selectivity of MSTd neurons for radial and rotational motion reflected the spatial pattern of translations present in these stimuli. Furthermore, using a set of eight flow stimuli (expansion, contraction, clockwise rotation (CW), counterclockwise rotation (CCW), and four intermediate spiral patterns), Graziano et al. (1994) found that most MSTd cells responded to more than one of these stimuli and often preferred intermediate directions of flow not corresponding to cardinal directions of spiral motion (i.e., pure expansion, contraction, or rotation). Interestingly, the study found that 57 of 66 of MSTd cells (86%) had a Gaussian tuning curve across the range of all rotational 2D flow fields ("spiral space"; $\sigma = 1$ and $\phi \in [0, 2\pi)$ in (1)).

We simulated the experiment from Graziano et al. (1994) by measuring the responses of MSTd-like model units to full-field spiral stimuli. In order to generate the receptive fields of model MSTd, we performed NMF with $B = 16$ on a set of 1500 flow fields sampled from the full range of 2D flow fields with translational, rotational, and deformational components (see Section "Optic flow stimuli"), yielding 16 MSTd-like model units. In order to facilitate statistical comparisons between model responses and biological data, NMF with $B = 16$ was repeated four times with different random initial conditions to generate a total of 64 MSTd-like model units. As in the study of Graziano et al. (1994), we fit the resulting tuning curves with a Gaussian function to find the peak (the mean of the Gaussian) that corresponded to the preferred direction in spiral space, and to provide a measure of bandwidth ($\sigma$, the SD of the Gaussian) and goodness-of-fit ($r$, the correlation coefficient).

Example receptive fields of MSTd-like model units are shown in Figure 2. Here, receptive fields are visualized as flow fields by calculating the population vector across the synaptic weight contributions of MT-like model units coding for a particular spatial location (Georgopoulos et al., 1982). The response of model MSTd to different input stimuli is indicated with open circles. Solid lines are the Gaussian fit, and the vertical lines indicate the peak response (mean of the Gaussian). Consistent with Graziano et al. (1994), receptive fields of model units had preferred directions that covered the full range of spiral motions (A–G), including flow fields that could be easily thought of as heading templates (A, G, and even B). Similarly, although displaying smooth Gaussian tuning in spiral space, some model units could be thought of as planar units (C and E) if their receptive fields are not probed across their full spatial extent. Figure 2H shows an example of a unit whose tuning curve was not well fit with a Gaussian function. This unit appeared to have mostly planar tuning.

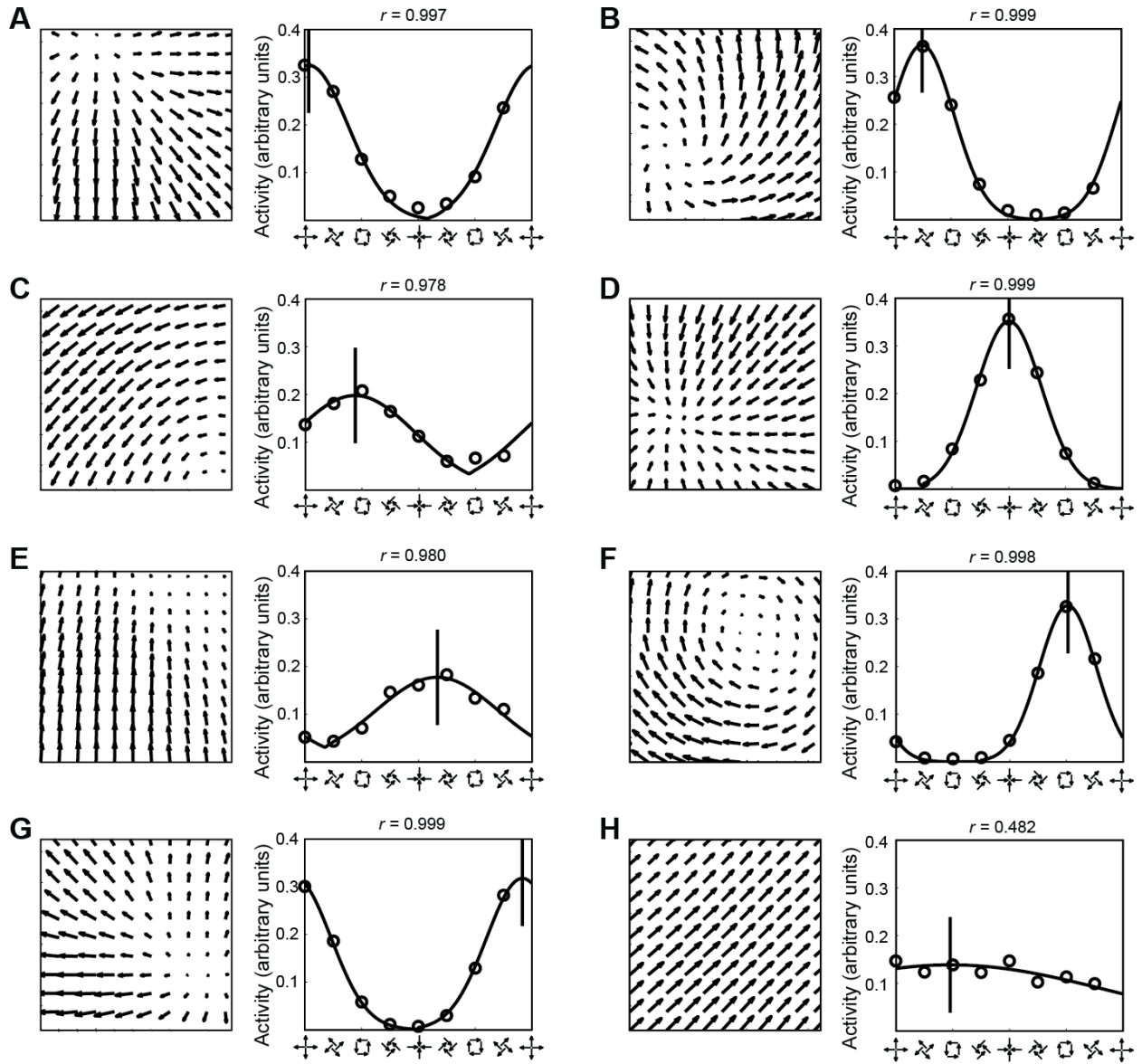

Figure 2: Example receptive fields of MSTd-like model units, visualized as flow fields by calculating the population vector across the synaptic weight contributions of MT-like model units coding for a particular spatial location. Open circles are the model response to different input stimuli. Solid lines are the Gaussian fit, and the vertical lines indicate the peak response (mean of the Gaussian). Receptive fields covered the full range of spiral motions (A–G), including flow fields that could be mistaken for heading templates (A, G, and even B). Some tuning curves were not well-fit with a Gaussian (H).

When looking at the preferred motion direction of MSTd cells (Figure 3A), Graziano et al. (1994) observed Gaussian tuning across the full range of rotational flow fields. Here, 0° corresponded to clockwise rotation, 90° to expansion, 180° to counterclockwise rotation, 270° to contraction, and the oblique directions (45°, 135°, 225°, and 315°) corresponded to four intermediate spiral stimuli. Each arrow indicated the preferred direction or peak response (the mean of the Gaussian) of each neuron ($N = 57$) in spiral space. Interestingly, their study revealed a strong bias toward expanding flow stimuli (Figure 3C), and found that stimuli progressively farther from expansion were progressively less represented (Graziano et al., 1994). This bias had been previously reported by Tanaka and Saito (1989). Graziano et al. (1994) concluded that MSTd did not use an axis-based method to analyze motion, but instead sampled a continuous array of stimuli with a strong bias for the expanding stimulus.

We used the bias for expanding stimuli that was observed empirically to match the population data from Graziano et al. (1994). This was achieved by sampling the offset, $\phi$, of all input stimuli according to the empirically observed likelihood of preferred directions of motion (Graziano et al., 1994; Clifford et al., 1999) (see Section "Optic flow stimuli"). The results are shown in Figure 3B and Figure 3D. Consistent with their data, the majority of preferred spiral directions clustered near expansion, with only few units preferring contraction, and roughly 47% of units were spiral-tuned (vs. 35% in their sample). Similar to 86% of isolated MSTd cells having smooth tuning curves and good Gaussian fits (i.e., a correlation coefficient of $r \geq 0.9$, with an average $r$ of 0.97), 43 of 64 MSTd-like model units (67%) had $r \geq 0.9$, with an average $r$ of 0.972. Compared to real MSTd neurons, the model units had comparable, although slightly broader Gaussian widths (an average $\sigma$ of 89.2° and SE of 13.6° in our case versus an average $\sigma$ of 61° and SE of 5.9° in their case).

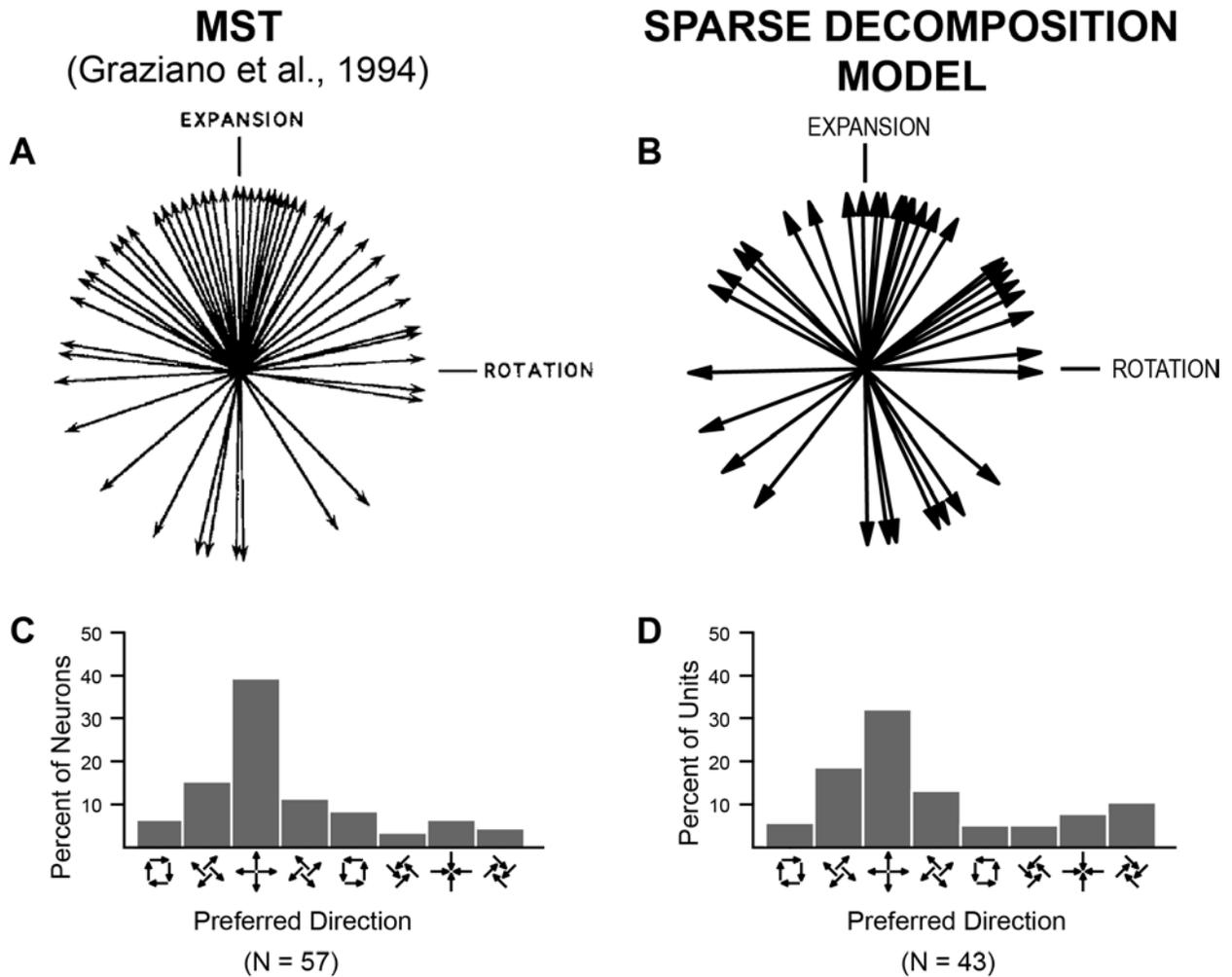

Figure 3: Gaussian tuning in spiral space. A, Gaussian tuning of a sample of 57 neurons across the full range of rotational flow fields, reprinted from Graziano et al. (1994) (their Fig. 9). Each arrow indicates the peak response (the mean of the Gaussian fit) of each neuron in spiral space. B, The distribution of preferred spiral directions of a sample of 43 MSTd-like model units whose tuning curves were well-fit with a Gaussian. C, Bar plot of the data in A, for better comparison, adapted from Clifford et al. (1999). D, Bar plot of the data in B.

Similar measures were achieved by the template model (Perrone and Stone, 1998) when it was applied to flow fields that simulated forward motion over a ground plane ("ground configuration"), although spiral tuning was slightly less frequent in their sample (27% in the ground configuration), and tuning curves were slightly sharper (76% of heading templates had $r > 0.9$ with a mean $\sigma$ of 52°).

## 3.2 Continuum of response selectivity

A continuum of response selectivity in MSTd was first reported by Duffy and Wurtz (1991a, 1995). Using a "canonical" set of twelve flow stimuli (eight directions of planar motion; expansion and contraction; clockwise and counterclockwise rotation) they showed that most neurons in MSTd were sensitive to multiple flow components, with only few neurons responding exclusively to either planar, circular, or radial motion. These results were independent of the eccentricity of the receptive field and the exact spatial location of the center of motion (COM). In a sample of 268 MSTd cells, Duffy and Wurtz (1995) found that 18% of cells primarily responded to one component of motion (planar, circular, or radial), that 29% responded to two components (planocircular or planoradial, but rarely circuloradial), and that 39% responded to all three components (Figure 4A).

We simulated their experiments by presenting the same twelve stimuli to a population of 64 MSTd-like model units and classified their responses. Duffy and Wurtz (1995) classified neurons according to the statistical significance of their responses, which was difficult to simulate since MSTd-like model units did not have a defined noise level or baseline output. Instead we mimicked their selection criterion by following a procedure from Perrone and Stone (1998), where the response of a model unit was deemed "significant" if it exceeded 12% of the largest response that was observed for any model unit in response to any of the tested stimuli.

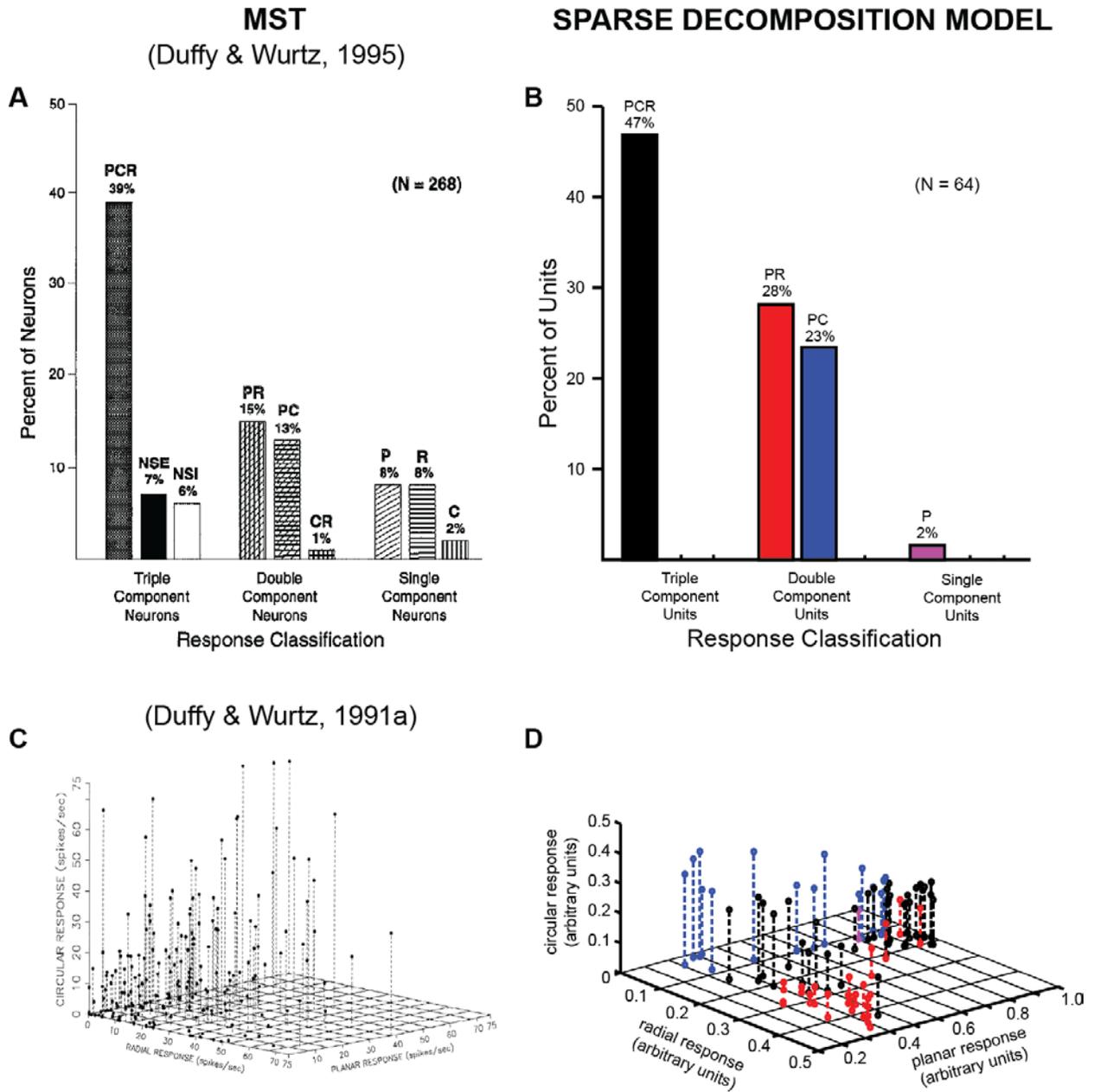

**Figure 4: Responses of all neurons and model units according to their responses to one, two, or three types of visual motion. A,** Response classification of a sample of 268 MSTd neurons as reported in Fig. 3 by Duffy and Wurtz (1995). **B,** Response classification of 64 MSTd-like model units. Triple-component cells are: planocirculoradial (PCR), nonselective excitatory (NSE), and nonselective inhibitory (NSI). Double-component cells are: planoradial (PR), planocircular (PC), and circuloradial (CR). Single-component cells are: planar (P), radial (R), and circular (C). 81% of neurons in A and 98% of model units in C respond to more than one type of motion. **C,** Distribution of a

sample of 220 MSTd neurons according to their responses to planar, radial, and circular stimuli, as reported in Fig. 8 by Duffy and Wurtz (1991a). D, Distribution of the 64 MSTd-like model units shown in B. Neurons and model units vary continuously in their responses to planar, radial, and circular stimuli.

Without further tuning, our model recovered a distribution of response selectivities very similar to those reported by Duffy and Wurtz (1995) (Figure 4B): The largest group was formed by triple-component or planocirculoradial (PCR) MSTd-like model units with selective responses to planar, circular, and radial stimuli (black). Double-component units were mainly planoradial (PR) with responses to a planar or radial stimulus (red) and planocircular (PC) units with responses to a planar or circular stimulus (blue). In contrast to Duffy and Wurtz (1995), we did not find any nonselective excitatory (NSE), radial (R) or circular (C) units, and only few planar (P) units (magenta). Our model did not include inhibitory units, and thus could not recover any nonselective inhibitory (NSI) responses.

In their earlier study, Duffy and Wurtz always positioned the stimuli on the receptive field of the neurons (Duffy and Wurtz, 1991a), whereas in their later study they always placed the stimuli on the center of the screen (center of gaze) (Duffy and Wurtz, 1995). They achieved similar results for both configurations. Similarly, all our stimuli were placed in the center of the screen and receptive field, without regard to the preferred COM location of a unit's receptive field.

Only 1% of the cells observed by Duffy and Wurtz (1995) were circuloradial (CR) cells, which they suggested were equivalent to the spiral-selective neurons reported by Graziano et al. (1994). Thus they concluded that spiral tuning was rare in MSTd. Interestingly, our model recovers distributions that are comparable to both empirical studies; that is, an abundance of spiral-tuned cells in Figure 3, and an abundance of PCR cells in Figure 4. Our results thus offer an alternative explanation to these seemingly contradictive findings, by considering that most spiral-tuned cells, as identified by Graziano et al. (1994), might significantly respond to planar stimuli when analyzed under the experimental protocol of Duffy and Wurtz (1995), effectively making most spiral-tuned cells part of the PCR class of cells, as opposed to the CR class of cells.

Duffy and Wurtz (1991a) also observed that single-, double-, and triple-component neurons showed substantial variation in the strength of their responses. This is illustrated in a 3D plot in which the responses of each neuron are represented as a single dot (Figure 4C), where the location of each dot along the planar, circular, and radial axes represents the response amplitude to that motion component. Combining single-, double-, and triple-component neurons revealed that the variation in component strength was continuous throughout the sample, and that there was no indication of significant clustering into separate response groups.

Similarly, our model recovered a continuum of response selectivity (Figure 4D). Here, the classification of each neuron is color-coded according to the colors used in Figure 4B. Consistent with data from Duffy and Wurtz (1991a), model units could not be categorized into qualitatively distinct response classes, instead displaying significant overlap among color groupings. Similar to biological MSTd, mild clustering can be observed for PC and PR neurons because, by definition, PC model units have a small R component, and PR model units have a small C component. However, due to the scarcity of single-component units, only few points were clustered along the cardinal axes. Overall model units seemed to be overly sensitive to planar components of motion (note the different scale on the planar response axis). This effect could be easily counteracted by adding a nonlinear, potentially compressive (Mineault et al., 2012), term to the motion integration in (4).

### 3.3 Spatial distribution of preferred centers of motion (COMs)

When studying how the responses of neurons in macaque MSTd changed as a function of the COM location of optic-flow stimuli, Duffy and Wurtz (1995) noticed that a large fraction of neurons gave their strongest responses when the COM was shifted away from the center of the stimulus. They investigated whether there was any preference among the neurons under study for one part of the visual field, and observed three possible response patterns. Those with their strongest responses to: 1) centered (COM at 0° of the visual field), 2) eccentric (COM at 45° of the visual field), and 3) peripheral COMs (COM at 90° of the visual field).

Figure 5A replots the relative distribution of COMs that evoked the strongest response for MSTd neurons that preferred radial (expansive and contractive) flow stimuli (compare Fig. 9B in Duffy and Wurtz (1995)). Analogously, Figure 5C replots the relative distribution of COMs that evoked the strongest responses for neurons that preferred circular (clockwise and counterclockwise rotation) flow stimuli (compare Fig. 9D in Duffy and Wurtz (1995)).

## MST
(Duffy & Wurtz, 1995)

## SPARSE DECOMPOSITION MODEL

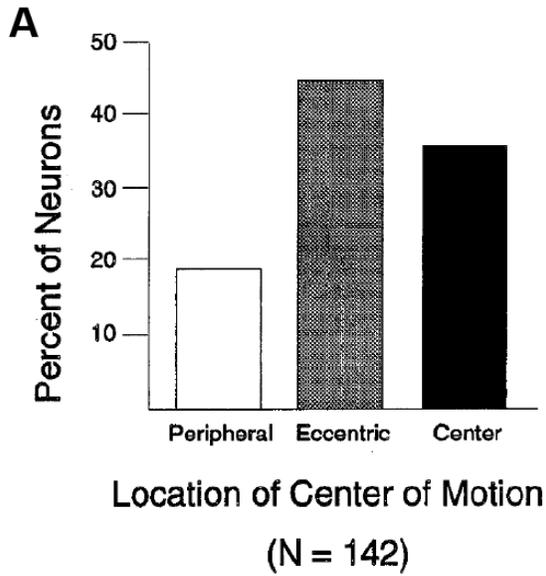

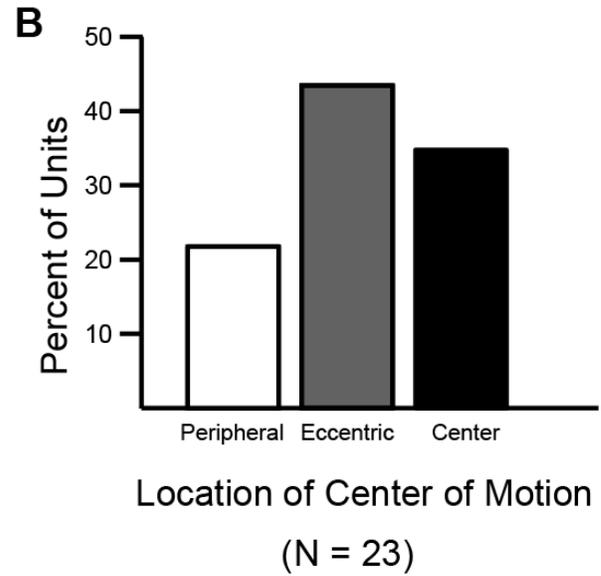

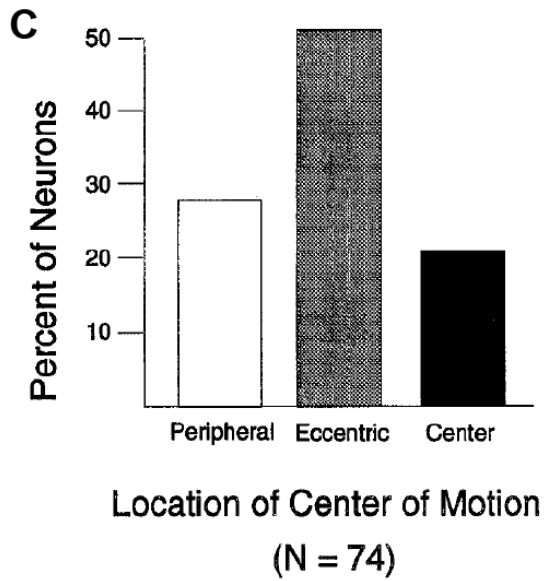

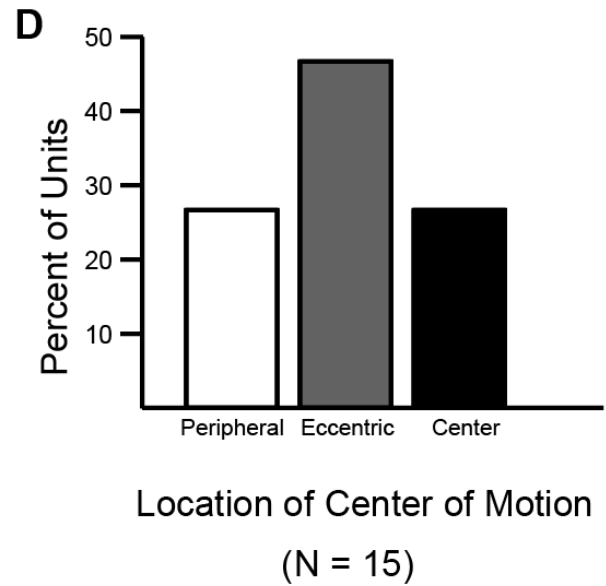

Figure 5: Spatial distribution of preferred centers of motion (COMs). A, The relative distribution of COMs that evoked the strongest response for a sample of 142 MSTd neurons that preferred radial (expansive and contractive) flow stimuli, as reported in Fig. 9B by Duffy and Wurtz (1995). B, The relative distribution of COMs for 23 MSTd-like model units that preferred radial flow stimuli. C, The relative distribution of COMs for a sample of 142 MSTd neurons that preferred circular (clockwise and counterclockwise rotation) flow stimuli, as reported in Fig. 9D by Duffy and Wurtz (1995). D, The relative distribution of COMs for 15 MSTd-like model units that preferred circular flow fields. Neurons and model units prefer eccentric stimuli, and the central area of the visual field appeared to be represented by a relatively greater proportion of units.

We mimicked their experiment by simulating the responses of MSTd-like model units to full-field stimuli with shifted COMs. We restricted our simulations to neurons whose Gaussian curves had a goodness-of-fit of $r \geq 0.9$, and determined their preferred direction of motion in spiral space (see Section "Gaussian tuning in spiral space"). We then generated stimuli with the same preferred direction and COMs shifted across all possible spatial locations. We defined the center area of the 32 × 32 pixel stimuli to have a radius of 6 pixels ($\approx 14°$), the eccentric area to lie outside the center area with a maximal radius of 11 pixels ($\approx 27°$), and the peripheral area to lie outside both the center and eccentric area with a maximal radius of 16 pixels ($\approx 40°$). We found very similar distributions to the findings by Duffy and Wurtz (1995) for model units selective for radial ($\phi \in \left\{\left[-\frac{\pi}{4}, \frac{\pi}{4}\right), \left[\frac{3\pi}{4}, \frac{5\pi}{4}\right)\right\}$) stimuli (Figure 5B) and for units selective for circular ($\phi \in \left\{\left[\frac{\pi}{4}, \frac{3\pi}{4}\right), \left[\frac{5\pi}{4}, \frac{7\pi}{4}\right)\right\}$) stimuli (Figure 5D), even though we included a wider range of stimuli and stimulus selectivity than Duffy and Wurtz (1995). Consistent with their data, we found that most model units preferred eccentric COMs and that the central area of the visual field appeared to be represented by a relatively greater proportion of units (Duffy and Wurtz, 1995).

### 3.4 Position invariance

Studies have reported that some MSTd neurons maintain their preference for a particular motion pattern even when the COM is shifted across the receptive field (Duffy and Wurtz, 1991b; Graziano et al., 1994; Lagae et al., 1994). This finding led researchers to suggest that some neurons in MSTd might be selective for the global pattern of motion in a way that could not be explained by local directional selectivity (Graziano et al., 1994), considering that for certain stimuli, such as rotational flow, the local direction of motion can change (or even reverse) depending on the spatial location.

For a sample of MSTd cells, Graziano et al. (1994) presented small ($\approx 10°$ diameter) stimuli at five positions in the receptive field, where one position was the center of the RF and the other four positions were arranged in an overlapping cloverleaf arrangement equally distributed around the RF center. To visualize the resulting neuronal responses, they devised a position invariance index (PI). For each cell, they calculated the directional selectivity (DS) at each of the five positions by comparing the response to the preferred stimulus (e.g., expansion) to the response to the antipreferred stimulus (e.g., contraction) [DS $= 1 -$ (response to antipreferred stimulus/response to preferred stimulus)]. The DS at each of the four surrounding position was then divided by the DS at the central position (PI $=$ $DS_{surround}/DS_{center}$), yielding four PI values per cell. A perfectly position-invariant cell, responding equally to all tested locations, would have all PI indices equal to 1; a cell that responded slightly differently depending on the location would have PI indices that varied around 1; and a cell that reversed its selectivity in some locations would have negative PI indices. Interestingly, they reported a sharp peak centered at PI $= 1$, indicating that most cells in their sample ($N = 52$) were perfectly position-invariant. Also, none of the PI values dropped below 0, indicating that no cells reversed their selectivity.

We simulated their experiments by calculating PI indices for a sample of MSTd-like model units using small-field stimuli (Figure 6A, B). We again restricted our simulations to neurons whose Gaussian curves had a goodness-of-fit of $r \geq 0.9$, and determined their preferred direction of motion in spiral space (see Section "Gaussian tuning in spiral space"), but in contrast to Graziano et al. (1994) also included spiral-tuned model units. We then generated stimuli with the same preferred direction and COMs in one of four possible locations, arranged in a cloverleaf arrangement, and calculated the model unit's PI index. Figure 6A shows the four presented small-field stimuli (with a diameter of 5 pixels or roughly 12.5°) for an example MSTd-like model unit (inset), with the corresponding PI value indicated above each stimulus. For the sake of completeness we also plot the center stimulus, which had PI = 1 per definition. The further away the COM of the presented stimulus was shifted from the preferred COM of the unit, the lower the PI value. Because the unit's preferred COM was not in the center of the stimulus, PI values larger than 1 were found.

Repeating this procedure for all model units in the sample (but excluding the PI value of center stimuli) revealed a distribution sharply peaked around PI = 0.95 (Figure 6B), indicating almost perfect position invariance, with a range of PI values that was very similar to those reported by Graziano et al. (1994). However, similar to Perrone and Stone (1998), we found a few negative PI values indicating that some model units occasionally changed their directional preference. All of these negative PI values were encountered for model units whose preferred COM was far away from the center of the stimulus, resulting in a small value for $DS_{center}$.

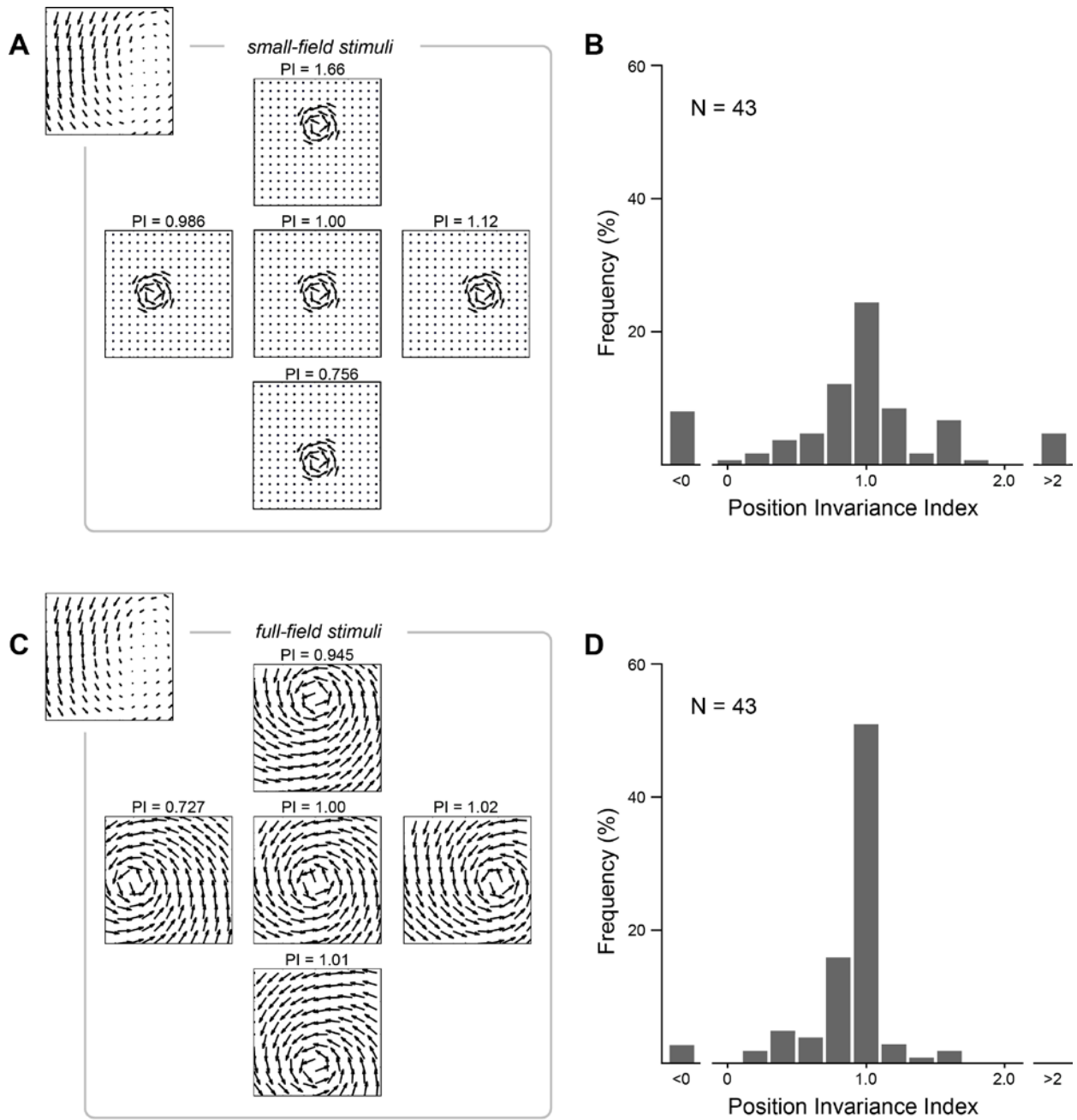

**Figure 6: Position invariance test. A**, Example model unit tested with small-field stimuli. On each trial, the stimulus appeared at one of five stimulus locations within the receptive field. **B**, Resulting distribution of position invariance (PI) index values for all 43 model units tested with small-field stimuli. PI=1 indicates perfect position invariance. Negative PI values indicate that the unit reversed its stimulus preference depending on the location of the stimulus. **C**, Example model unit with full-field stimuli. **D**, Resulting distribution of PI values for all 43 model

units tested with full-field stimuli. PI values were tightly clustered around PI=1 for both small-field and full-field stimuli, and only a few negative PI values were found.

The effect of larger moving stimuli over larger distances has also been examined (Duffy and Wurtz, 1991a; Lagae et al., 1994; Duffy and Wurtz, 1995). These procedures generally revealed larger variations in the response amplitudes depending on stimulus location, because it was possible for stimuli to fall only partially within the receptive field of a neuron. However, when focusing on the binary directional preference along a cardinal axis of motion (e.g., expansion versus contraction), Duffy and Wurtz (1991a, 1995) found that most MSTd neurons retained their directional preference over their entire receptive field.

Compared to data from Duffy and Wurtz (1991a, 1995), repeating the simulation for full-field stimuli (Figure 6C, D) revealed a similar distribution of PI values. In fact, the full-field distribution was even sharper than the one obtained with small-field stimuli (Figure 6A, B): PI values were still tightly clustered around PI = 0.96, indicating almost perfect position invariance, but no longer exceeded a value of 2, and only a few negative PI values remained.

Therefore, consistent with biological MSTd (Duffy and Wurtz, 1991a; Graziano et al., 1994; Lagae et al., 1994; Duffy and Wurtz, 1995), MSTd-like model units exhibited limited position invariance (defined as a response that maintains its directional preference) when tested with either small-field or full-field stimuli.

## 3.5 Representation of heading using a sparse population code

During forward movement, retinal flow radiates out symmetrically from a single point, the focus of expansion (FOE), from which heading (i.e., the instantaneous direction of travel) can be inferred (Gibson, 1950). Studies have shown that heading can be decoded to a very high degree of precision from the population response of neurons in MSTd. Page and Duffy (1999) found that the position of the FOE could be decoded within $\pm 10°$ of precision from one second of MSTd spike trains. Ben Hamed et al. (2002) then went on to show that the FOE in both eye-centered and head-centered coordinates could be decoded from MSTd population activity with an optimal linear estimator even on a single-trial basis, with an error of $0.5 - 1.5°$ and SD of $2.4 - 3°$.

Therefore, in order to assess whether model MSTd could make accurate predictions about heading, we tested whether the 2D coordinates $(x_{FOE}, y_{FOE})$ of the FOE of arbitrary expansive flow fields could be decoded from a population of $N$ MSTd-like model units using a set of $N \times 2$ linear weights and $1 \times 2$ bias values. Input stimuli were a set of $10^3$ pure expansive flow fields (having $\sigma = 1$, $\phi = 0$ in (1)) with randomly selected FOEs, for which we calculated model MSTd output. We performed five-fold cross-validation with a training set containing 800 stimuli and a test set containing 200 stimuli. The model was trained for 50 epochs on the training set using gradient descent on the Euclidean distance (mean squared error) between predicted and actual FOE coordinates, with a learning rate of $10^{-1}$, momentum 0.9, and batch size 16. The experiment was repeated for a number of basis vectors $B = 2^i$, where $i \in \{2, 3, ..., 7\}$.

The result is shown in Figure 7A. Each data point shows the model's heading prediction error on the test set for a particular number of basis vectors, averaged over five trials (i.e., the five folds of the cross-validation procedure). The vertical bars are the SD. Model MSTd accurately encoded heading, often with sub-pixel accuracy. The smallest heading prediction error of $0.312 \pm 0.24$ pixels (roughly corresponding to $0.779° \pm 0.60°$) was achieved with $B = 16$. Note that these numbers were achieved on the test set; that is, on flow fields the model had never seen before. Training error was zero in all cases. Heading prediction error on the test set thus corresponded to the model's ability to generalize; that is, to deduce the appropriate response to a novel stimulus using what it had learned from other previously encountered stimuli (Hastie et al., 2009; Spanne and Jorntell, 2015).

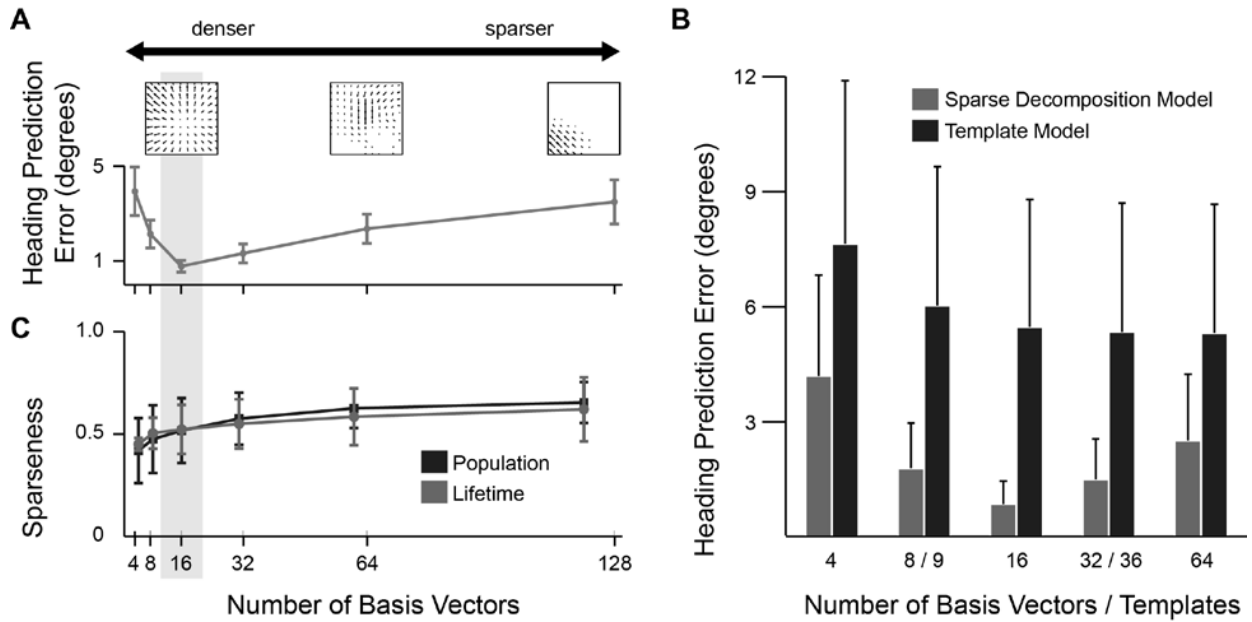

Figure 7: Sparse population coding of heading. A, Heading prediction (generalization) error as a function of the number of basis vectors using five-fold cross-validation. Vertical bars are the SD. B, Quantitative comparison of heading prediction (generalization) error for the sparse decomposition model, compared to the template model with the same number of templates. The sparse decomposition model consistently and significantly (Kolmogorov-Smirnov test, $p < 〚10〛^{(-4)}$) outperformed the template model. C, Population and lifetime sparseness as a function of the number of basis vectors. Operating the sparse decomposition model with B=16 basis vectors co-optimizes for both accuracy and efficiency of the encoding, and leads to basis vectors that resemble MSTd receptive fields.

In order to compare our results to the template model (Perrone and Stone, 1994, 1998), we repeated the experiment for a set of $N$ heading templates, with their preferred ($x_{\text{FOE}}, y_{\text{FOE}}$) equally distributed across the visual field. In the template model, heading is usually inferred by the preferred FOE of the most active heading template. However, this approach would require on the order of $10^3$ heading templates to achieve heading prediction errors on the order of 1 pixel (or 1°) (Perrone and Stone, 1998). Therefore, in order to allow for a fair comparison, we decoded heading using the same experimental protocol using a set of $N \times 2$ linear weights and $1 \times 2$ bias values. Still, the sparse decomposition model consistently and significantly (Kolmogorov-Smirnov test, $p < 10^{-4}$) outperformed the template model with an equal number of heading templates (Figure 7B).

We next asked how the number of basis vectors influenced the sparseness of the population code (Figure 7C) by investigating how many MSTd-like model units were activated by any given stimulus (population sparseness) as well as how many stimuli any given model unit responded to (lifetime sparseness). We applied the model to a set of $10^3$ randomly chosen flow fields across the full parameter range in Equation (1), and computed sparseness metrics on the resulting population responses. Sparseness metrics were computed according to the definition by Vinje and Gallant (2000), which can be understood as a measure of both the nonuniformity ("peakiness") and strength of the population response (see Section "Sparseness"). A sparseness value of zero would be indicative of a dense code (where every stimulus would activate every neuron), whereas a sparseness value of one would be indicative of a local code (where every stimulus would activate only one neuron) (Spanne and Jorntell, 2015). As expected, the analysis revealed that both population and lifetime sparseness monotonously increased with an increasing number of basis vectors. Sparseness values ranging from roughly 0.4 for $B = 4$ to 0.65 for $B = 128$ suggested that the population code was sparse in all cases, as it did not get close to the extreme case of either a local or a dense code.

Overall these results suggest that model MSTd operates in a sparseness regime that is well-suited to efficiently encode perceptual variables such as heading (Ben Hamed et al., 2003). Interestingly, we found that the sparseness regime in which model MSTd achieved the lowest heading prediction error (Figure 7C) was also the regime in which recovered basis vectors most resembled receptive fields of macaque MSTd (Figures 2–6).

## 4. Discussion

We found that applying NMF (Paatero and Tapper, 1994; Lee and Seung, 1999, 2001) to MT-like patterns of activity can account for several essential response properties of MSTd neurons, such as tuning to radial, circular, and spiral motion patterns (Duffy and Wurtz, 1991a, b; Graziano et al., 1994; Lagae et al., 1994; Duffy and Wurtz, 1995), heading selectivity (Page and Duffy, 1999; Ben Hamed et al., 2003; Page and Duffy, 2003), and limited position invariance (Graziano et al., 1994; Duffy and Wurtz, 1995). This finding suggests that these properties might emerge from MSTd neurons performing a biological equivalent of dimensionality reduction on their inputs. At the population level, model MSTd efficiently and accurately predicts heading using a sparse distributed code (Olshausen and Field, 1996, 1997; Ben Hamed et al., 2003), consistent with ideas from the efficient-coding and free-energy principles (Attneave, 1954; Barlow, 1961; Linsker, 1990; Simoncelli and Olshausen, 2001; Friston et al., 2006; Friston, 2010). The present work provides an alternative to the template-model view of MSTd (Perrone and Stone, 1994, 1998), and offers a biologically plausible account of the receptive field structure across a wide range of visual response properties in MSTd.

## 4.1 Sparse decomposition model of MSTd

It is well-known that MSTd neurons do not decompose optic flow into its first-order differential invariants (i.e., into components of divergence, curl, and deformation), because: 1) a significant amount of MSTd neurons respond best to spiral motion (Graziano et al., 1994), 2) only few neurons are selective for deformation (Lagae et al., 1994), and 3) the response of an MSTd neuron decreases when its preferred component of flow is mixed with increasing amounts of a different component (Orban et al., 1992).

Here we provide computational evidence that MSTd neurons might decompose optic flow in the sense of matrix factorization, such that the spectrum of retinal flow fields encountered during self-movement can be represented in an efficient and parsimonious fashion. Such a representation is in agreement with the efficient-coding or infomax principle (Attneave, 1954; Barlow, 1961; Linsker, 1990; Simoncelli and Olshausen, 2001), which posits that sensory systems should employ knowledge about statistical regularities of their input in order to maximize information transfer. If information maximization is interpreted as optimizing for both accuracy (prediction error) and efficiency (model complexity), then the efficient-coding principle can be understood as a special case of the free-energy principle (Friston et al., 2006; Friston, 2010).

A sparse, parts-based decomposition model of MSTd implements these principles in that it can co-optimize accuracy and efficiency by representing high-dimensional data with a relatively small set of highly informative variables. Efficiency is achieved through sparseness, which is a direct result of NMF's nonnegativity constraints (Lee and Seung, 1999). Accuracy trades off with efficiency, and can be achieved by both minimizing reconstruction error and controlling for model complexity (i.e., by tuning the number of basis vectors) (see Figure 7). Statistical knowledge enters the model via relative frequencies of observations in the input data (e.g., manifested as a bias towards expanding flow fields). NMF explicitly discourages statistically inefficient representations, because strongly accounting for a rare observation at the expense of ignoring a more common stimulus component would result in an increased reconstruction error.

## 4.2 Representation of heading using a sparse population code

Accurate heading representations emerged in our population of MSTd-like model units. Consistent with single-neuron data from macaque MSTd (Tanaka and Saito, 1989; Duffy and Wurtz, 1991a, 1995; Gu et al., 2006), our model recovers individual MSTd-like model units that are selective to the expanding radial motion that occurs as an observer moves through the environment (Figure 2A, B, and G). However, our model is agnostic to the functional role that these neurons might play in perception and behavior. Adoption of the sparse decomposition model supports the view that these single-unit preferences emerge from the pressure to find an efficient representation of large-field motion stimuli, as opposed to encoding a single perceptual variable such as heading. Yet heading could be accurately inferred from the population activity of model MSTd with a bias on the order of $0.1° - 1°$, consistent with neurophysiological data (Ben Hamed et al., 2003; Page and Duffy, 2003).

Interestingly, we found that the sparseness regime in which model MSTd achieved the lowest heading prediction error and thus showed the greatest potential for generalization (Figure 7A, C) was also the regime in which recovered basis vectors most resembled receptive fields of macaque MSTd (Figures 2–6). In contrast to findings about early sensory areas (Olshausen and Field, 1996, 1997; Vinje and Gallant, 2000), this regime does not utilize an overcomplete dictionary, yet can still be considered a sparse coding regime (Spanne and Jorntell, 2015). Sparse codes are a trade-off between dense codes (where every neuron is involved in every context, leading to great memory capacity but suffering from cross-talk among neurons) and local codes (where there is no interference but also no capacity for generalization) (Spanne and Jorntell, 2015). We speculate that sparse codes with a relatively small dictionary akin to the one described in this paper might be better suited (as opposed to overcomplete basis sets) for areas such as MSTd, because the increased memory capacity of such a code might lead to compact and multi-faceted encodings of various perceptual variables (Bremmer et al., 1998; Ben Hamed et al., 2003; Brostek et al., 2014).

### 4.3 Model alternatives

Several computational models have tried to account for heading-selective cells in area MSTd. In the template model (Perrone and Stone, 1994, 1998), MSTd-like units receive input from a mosaic of MT-like motion sensors that correspond to the flow that would arise from a particular heading. A complication of this type of model is that it requires an extremely large number of templates to cover the combinatorial explosion of heading parameters, eye rotations, and scene layouts (Perrone and Stone, 1994), even when the input stimulus space is restricted to gaze-stabilized flow fields (Perrone and Stone, 1998). The velocity gain field model (Beintema and van den Berg, 1998, 2000) tries to reduce the number of combinations by using templates that uniquely pick up the rotational component of flow, but recent evidence argues against this type of model (Duijnhouwer et al., 2013).

Furthermore, these models do not address the fact that MSTd neurons exhibit a continuum of visual response selectivity to large-field motion stimuli (Duffy and Wurtz, 1991a; Graziano et al., 1994; Lagae et al., 1994; Duffy and Wurtz, 1995). An exception is the recent study by Mineault et al. (2012), which demonstrated that the "complex motion" preference of individual MSTd cells could be predicted by nonlinearly interacting RF subfields. However, the seemingly arbitrary substructure of RFs did not allow for any overarching organizational principle to be revealed. Our model differs in that it can explain a range of MSTd response properties by restricting the integration of feed-forward MT-like input to be linear. In a related approach to the one presented in this paper, Zemel and Sejnowski (1998) proposed that neurons in MST encode hidden causes of optic flow, by using a sparse distributed representation that facilitates image segmentation and object- or self-motion estimation by downstream read-out neurons. Although many of their hidden units showed spiral tuning, they did not discuss the neurophysiological plausibility of the read-out mechanism or quantitatively assess many of the response properties described in this paper.

### 4.4 Model limitations and future directions

Despite its simplicity the present model is able to explain a variety of MSTd visual response properties. However, a number of issues remain to be addressed in the future, such as the fact that neurons in MSTd are also driven by vestibular (Page and Duffy, 2003; Gu et al., 2006; Takahashi et al., 2007) and eye movement-related signals (Komatsu and Wurtz, 1988;

Newsome et al., 1988; Bradley et al., 1996; Page and Duffy, 1999; Morris et al., 2012). In terms of visual response properties, the present model does not incorporate speed tuning (Tanaka and Saito, 1989; Orban et al., 1995) or the spatial distribution and variability of receptive fields (Duffy and Wurtz, 1991a; Raiguel et al., 1997). However, the model could be readily extended to account for these effects.

At a population level, MSTd exhibits a range of tuning behaviors from pure retinal to head-centered stimulus velocity coding (Ben Hamed et al., 2003; Yu et al., 2010; Brostek et al., 2014) that include intermediate reference frames (Fetsch et al., 2007). From a theoretical standpoint, the sparse decomposition model seems a good candidate to find an efficient, reference frame-agnostic representation of various perceptual variables (Pouget and Sejnowski, 1997; Pouget and Snyder, 2000; Ben Hamed et al., 2003), but future iterations of the model will have to demonstrate the practicality of this approach.

Finally, there is evidence that neurons in MSTd represent 3D self-movement (Gu et al., 2006; Logan and Duffy, 2006) and that some neurons are selective not just for heading, but also for path and place (Froehler and Duffy, 2002; Page et al., 2015). Future iterations of the model will address these issues step-by-step.

## 5. Acknowledgments

Supported by the National Science Foundation Award #1302125, Intel Corporation, and Northrop Grumman Aerospace Systems. The authors declare no competing financial interests.